\documentclass[final,3p,times,twocolumn]{elsarticle}

\usepackage{graphicx}
\usepackage{amssymb}
\usepackage{url}
\usepackage{xcolor}

\biboptions{sort&compress}

\journal{Nuclear Instruments and Methods, Section B}

\begin{document}

\begin{frontmatter}


\title{Electronic stopping cross sections of tungsten\\to swift ions and comparisons with models}



\author[usp]{Tiago F. Silva\fnref{myfootnote1}}
\fntext[myfootnote1]{Corresponding author. e-mail: tfsilva@usp.br}
\author[usp]{Arilson Silva}
\author[usp]{Cleber L. Rodrigues}
\author[usp]{Nemitala Added}
\author[usp]{Manfredo H. Tabacniks}%
\author[ipen]{\\Flávio Matias}
\author[ipen]{Helio Yoriyaz}
\author[ipen]{Julian Shorto}

\address[usp]{Instituto de Física da Universidade de São Paulo. Rua do Matão, 1371 - Cidade Universitária - São Paulo - Brazil.}
\address[ipen]{Instituto de Pesquisas Energéticas e Nucleares. Avenida Professor Lineu Prestes, 2242 - Cidade Universitária - São Paulo - Brazil.}

\begin{abstract}
Accurate stopping power data for tungsten is crucial for ion beam analysis (IBA) techniques applied to fusion-related materials. In this work, we present new experimental measurements of the stopping power of tungsten for protons and alpha particles, addressing key gaps in fundamental databases. Our results provide a densely spaced dataset, refining the practical uncertainty limits to approximately 1.5\% for protons and 2.0\% for alpha particles. We critically compare our findings with semi-empirical and theoretical models, evaluating their performance in describing the stopping power of tungsten for light projectiles. By improving the accuracy and reliability of stopping power data, we contribute to the enhancement of the applicability of ion-beam methods for characterizing tungsten in fusion-related research. These findings contribute to the refinement of semi-empirical models and support the ongoing efforts to develop more precise theoretical frameworks for ion-solid interactions in high-Z materials.
\end{abstract}

\begin{keyword}
Stopping power \sep Ion beam analysis \sep Tungsten \sep 


\end{keyword}

\end{frontmatter}


\section{INTRODUCTION}
\label{sec:intro}

Stopping power data for protons and helium in tungsten is of critical importance. Ion beam analysis of fusion-related materials heavily relies on the accuracy of stopping power models for this material \cite{mayer_ion_2019, rubel_accelerator_2023}, with significant implications for studies of material migration within reactor interiors and for deuterium and hydrogen profiling in retention and storage characterization \cite{brezinsek_plasmawall_2017, maier_deuterium_2019}.

Accurate stopping power data also drive theoretical models. Recent ab initio calculations of the stopping power for protons in materials have proven to be an efficient method for providing reasonably accurate data utilizing the electronic structure of constituent elements \cite{matias_efficient_2024}. Tungsten stands out as a prime candidate for benchmarking these models since data for such a heavy element can contribute to improving semi-empirical models.

The scarcity of data in the IAEA database \cite{IAEA_expdata, montanari_iaea_2024} poses challenges for evaluating existing models and limits the accuracy of IBA results in tungsten-based materials. For instance, only nine experimental datasets are currently available for each of protons and helium ions, with very few data available for heavy ions. By contrast, for copper (one of the most extensively studied materials) there are 47 experimental datasets for protons, 22 for helium ions, and data for 30 different heavy ions.

To address this issue, the IAEA has launched a coordinated research project (CRP-F11023) aimed at, among other things,  increasing the availability of high-quality experimental data of stopping cross-sections of tungsten to protons, helium, copper, and iodine. This initiative seeks to enhance the applicability of ion beam analysis (IBA) in tungsten studies, contributing to the advancement of fusion reactor development.

Some results were recently published by the Uppsala group as part of this CRP \cite{MORO20211}. Here, we present the results obtained by the IBA facility at the University of S\~ao Paulo, also under this CRP, alongside a critical analysis comparing existing data with semi-empirical and theoretical models. Our goal is to understand the discrepancies and limitations of the models and to propose ways to improve semi-empirical approaches.

Our data stand out by providing statistically robust evidence of high accuracy within an energy range suitable for Nuclear Reaction Analysis (NRA), the reference technique for profiling light elements in fusion-related materials research.

\section{METHODS}

    Our experiments were conducted in transmission mode to assess the stopping power to protons and in backscattering mode to assess the stopping power to helium ions. In the following, we describe the sample production and characterization, as well as the experimental procedures.

    \subsection{Sample preparation and characterization}

        A 3-$\mu$m nominal thick foil (commercially available from Goodfellow) was used in transmission experiments with the aim of stopping proton measurements. The average areal density was experimentally determined as 6.0178(25) mg/cm$^2$ by direct mass-area ratio. The mass was obtained by a high-precision scale (Sartorius balance, model SE2) and the area was obtained using a calibrated high-resolution optical scanner. Unfortunately, our results could not take advantage of this tight precision due to foil inhomogeneities. The adopted procedure of data analysis is detailed in the next section.
         
        The foil purity was evaluated as 95(1) at.\%. with a 5 at.\% carbon contamination detected using resonant proton scattering. See the next section for more details on this analysis that confirmed the nominal specification. 
        
        A sputter-deposited thin film of tungsten was used in backscattering experiments to determine the stopping to helium. The nominal thickness was 70 nm, with the experimentally determined areal atomic density at 483(5)$\times10^{15}$~at./cm$^2$ by proton scattering (148(2) $\mu$g/cm$^2$ or 76.5(8) nm). The low energy loss of 2.18 MeV protons (below ten keV) in this film allowed for the density determination using a thin film approximation (independent of energy-loss models). No carbon or oxygen contamination was observed in proton scattering experiments at their respective resonant energies. No inhomogeneities were detected in sequenced measurements.       
        
        Particle-induced X-ray emission (PIXE) measurements revealed no contaminants above the sensitivity of tens of parts per million in both samples. Thus, no other contaminant correction was considered necessary.

    \subsection{Experimental procedures and data processing}
    \label{sec:exp_dpr}
    
        Data processing for the transmission experiment followed the standards defined in a previous work \cite{moro_traceable_2016}. The procedure consists of partially shadowing the detector's field of view, and thus protons backscattered by a thin gold foil are recorded with a fraction of the protons going directly to the detector while the complementary fraction passes through the foil. The energy loss in the foil is then measured by the energy shift of both signals in the spectra.
    
\begin{figure}
            \centering
            \includegraphics[width=1\linewidth]{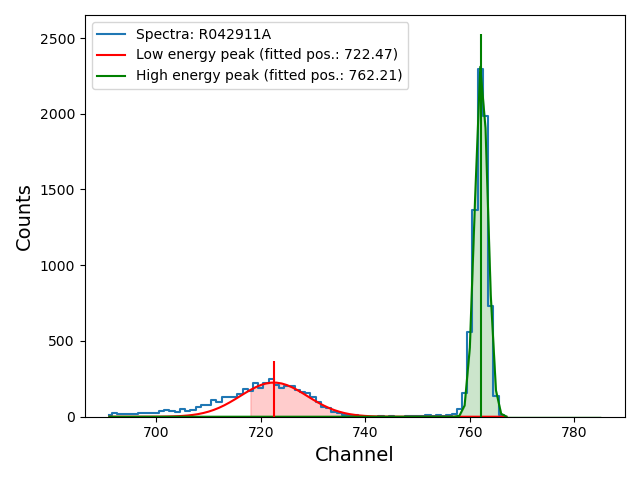}
            \caption{Typical data obtained in the transmission experiment. Shaded areas denotes the channels included in the Gaussian fit for the direct (right) and transmitted (left) peaks. Differences in the peak height is due to the restriction of assessed area in the stopping cross-section measurements.}
            \label{fig:transmission_data}
        \end{figure}

        A Gaussian curve is adjusted for both the direct and transmitted peaks. Firstly, we used a peak detection algorithm to obtain the initial value of the Gaussian center. Then a fine curve fit adjusts the centroid position and the width of the peaks. See Fig. \ref{fig:transmission_data}. Due to foil roughness, the transmitted peak presents a broader and asymmetric peak. To determine the peak position, only part of the data was used in the Gaussian fit. The data interval considered in the Gaussian peak is shown by the shaded regions in Fig. \ref{fig:transmission_data}. This procedure, inspired by other works \cite{mayer_skewness_2010, vomschee_note_2025}, is justified in terms of the determination of the most probable energy loss. 

\begin{figure}
            \centering
            \includegraphics[width=1\linewidth]{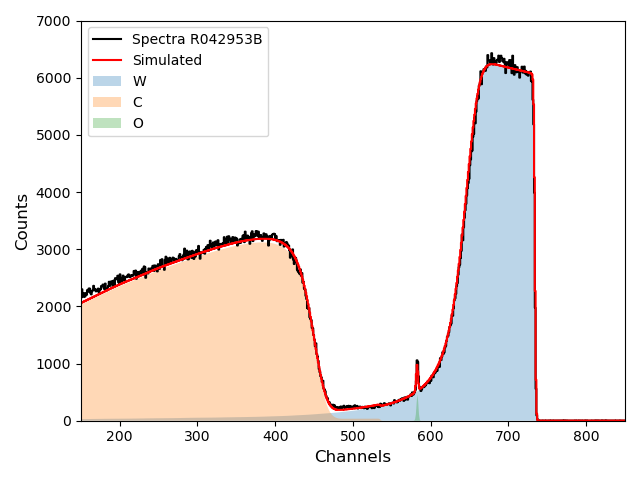}
            \caption{Proton scattering data used as constraint in the data processing method. This approach was used to compensate for sample inhomogeneities. The normally incident proton beam had 3600 keV kinetic energy and the detector was placed at 170$^\circ$ scattering angle. Carbon signal in the foil appear near the channel 500 and the small surface oxidation close to channel 600 was neglected.}
            \label{fig:proton_scattering}
        \end{figure}

        To address the inhomogeneities in the foil, we measured proton scattering spectra at the same position where the stopping cross-section measurements were performed (alignment precision better than 0.1 mm). A broad proton beam, covering an area comparable to that used in the stopping measurements (selected as approximately 1 mm$^2$), ensured that both measurements probed the same region of the sample, allowing for consistent comparison. Proton scattering measurements were conducted using a 3600 keV beam, and the detector placed at 170$^\circ$ scattering angle. The data are shown in Fg. \ref{fig:proton_scattering}.

        The data analysis followed an iterative approach. Starting from the areal density determined in the previous section, we calculated the stopping cross-section based on the measured energy loss, and then used this information as input for SIMNRA \cite{mayer_improved_2014} simulations of the scattering spectra (Fig. \ref{fig:proton_scattering}). The foil thickness was iteratively adjusted (and stopping cross-section updated accordingly) to achieve agreement between the experimental and the simulated scattering data. 

        This procedure converged to a foil areal density of 17.48(18)$\times$10$^{18}$ atoms/cm$^2$, corresponding to a thickness of 2.77(3) $\mu$m, which is slightly less than the nominal. The associated uncertainties were evaluated as 1\% using the MultiSIMNRA method, as described in \cite{silva_multisimnra:_2016}. The carbon contamination in the foil was obtained consistently in this iterative process.

        In backscattering experiments, the stopping information is extracted from the width of the tungsten peak in the spectra. For that, the sample was measured with a fixed detector at the 120$^\circ$ scattering angle under two incidence angle conditions, first at 0$^\circ$ and then at 60$^\circ$ referred to the normal vector to the surface.         


        Thus, computing the energy losses in the way-in and way-out of the film, assuming the surface approximation in both conditions, provides us with a system equation that reduces to:        
        \begin{equation}
            [\epsilon]_{in} = \frac{2 \cdot \Delta E_2 - \Delta E_1}{ 3 \cdot K \cdot t}
        \end{equation}
        and
        \begin{equation}
            [\epsilon]_{out} = \frac{2 \cdot \Delta E_1 - \Delta E_2}{3 \cdot t}
        \end{equation}
        with $\Delta E_1 = \Delta E(0^\circ,60^\circ)$ and $\Delta E_2 = \Delta E(60^\circ,0^\circ)$. These values are attributed to the average energies in the entrance and exit path of the ion in the film, calculated here as:
        \begin{equation}
        \label{eq:mean_energies}
            \bar{E}_{in} = E_0 - \frac{1}{4} \Delta \bar{E}
        \end{equation}
        
         \begin{equation}
            \bar{E}_{out} = K \cdot E_0 - \left( \frac{2 \cdot K + 1}{4} \right) \Delta \bar{E}
        \end{equation}
        with $\Delta \bar{E} = ( \Delta E_1 + \Delta E_2) /2 $. This method is described in detail in reference \cite{chu_chapter_1978}.

        \begin{figure}
            \centering
            \includegraphics[width=1\linewidth]{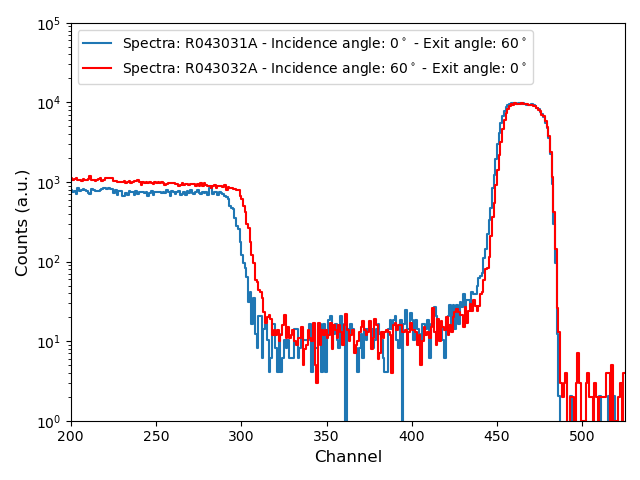}
            \caption{Typical data obtained in the scattering experiment. Differences in the tungsten peak width are due to different configuration of incidence and exit angles to the helium ions. The incidence energy was 2770 keV. Data obtained with incidence angle of 0$^\circ$ were scaled by a factor 2 so the tungsten peak presents the same height in the plot. }
            \label{fig:helium_scattering}
        \end{figure}
        
        Systematic uncertainties were estimated by propagating the film thickness and contaminant fraction uncertainties, while statistical uncertainties were estimated by propagating calibration and fit uncertainties together with a Monte Carlo approach using resampling of the spectra with Poisson statistics to propagate uncertainties related to counting statistics, as described in \cite{moro_traceable_2016}. The latter is used to evaluate the robustness for peak centroid and edges determination in transmission and backscattering experiments, respectively.   
    
    \subsection{Comparing models}
    
        We used for comparison two types of models. The semi-empirical ones, which are widely adopted for data processing in IBA, and theoretical models, mostly used to study ion-solid interactions. Each theoretical model has its own approach, and comparison with experimental data is always useful to constrain the models and provide feedback of model tailoring.
    
        \subsubsection{Semi-empirical models}
    
            The semi-empirical approaches are based on the adjustment of a theory-guided function to a selection of experimental data. Due to its wide acceptance in practical applications, comparisons with new and accurate experimental data are highly desirable. One key factor is that semi-empirical approaches tend to be accurate when reliable experimental data are available. For the case of tungsten, only nine experimental data sets are available in the IAEA database \cite{IAEA_expdata} for each proton and helium.
     
            In this work, we compared our results with the model contained in the ICRU Report 49 \cite{ICRU49}, which is adopted in the PSTAR and ASTAR databases hosted by NIST \cite{NIST_ST} and used as a reference in the Geant4 framework \cite{Geant4_2003, Geant4_2006, Geant4_2016}. We also compared with SRIM in its 2013 software version (whose last update of its database was reported in \cite{SRIM-2010}) since this is the most used database for ion implantation and simulations of defects creation on materials. Neither database has received updates for at least 10 years, but comparisons with new experimental data may provide insights into their quality, guiding improvements in practical applications.
    
            We also compared our results with the compilation of experimental data provided by the IAEA \cite{montanari_iaea_2024}. Only a few experimental data of stopping cross sections of tungsten to protons are contained in the database, with data from the 1970s and 1980s. For the He ions, the number of experimental results in the database is higher but far from abundant. In this case, the data are from the 1960s and early 1970s. In both cases, the Uppsala group added new experimental data in 2021. Data from this group were obtained in a wide energy range, significantly contributing to the characterization of the H and He energy loss processes in this material. However, additional data are needed to better assess systematic errors, and our work helps fill some gaps in the energy range covered by the Uppsala group.

        \subsubsection{Theoretical models}
    
            Theoretical models used in the comparison are: CasP \cite{casp_site}, DPASS \cite{dpass_site} and CBETHE \cite{Salvat2022, salvat_sbethe_2023}.

            CasP (Convolution Approximation for Swift Particles) is based on a Unitary Convolution Approximation (UCA) \cite{CasP2011}, where the energy loss of a swift ion is calculated by integrating the individual contributions of electrons. This method efficiently provides stopping power predictions over a broad energy range and for diverse ion-target combinations. CasP allows the replacement of UCA calculations for valence electrons by calculations based on the free-electron-gas approximations IDA (Induced Density Approximation) and TCS (Transport Cross-Section), as described in \cite{grande_alternative_2016, matias_ground-_2017}. Important information is that in this work calculations were performed with the option for scanning charge states of the ion, with the recombination, shell and Barkas corrections included in the model, and the number of valence electrons as 6. The stopping contribution due to projectile-electron loss (target-induced projectile ionization and excitation) and electron capture for charge exchange at charge-equilibrium conditions was also included in the model.

            In contrast, DPASS \cite{DPASS2019} is an extension of the PASS code (Particle Stopping Simulation), which is based on binary stopping theory - a refinement of Bohr’s classical model \cite{sigmund_binary_2000,sigmund_binary_2002}. This approach describes the interaction of a fast moving ion with target electrons, incorporating screening effects, charge-state evolution, and shell corrections \cite{DPASS2019}. The latest version of DPASS significantly expands its applicability, covering 92 ion species in 92 elemental targets over an energy range from 1 keV/u to 1 GeV/u, with direct comparisons to experimental data from the IAEA database. Additionally, DPASS allows for stopping power calculations in compound materials via Bragg additivity and accounts for deviations from this rule in selected cases.
    
            CBETHE \cite{Salvat2022} is a recent implementation of the Bethe stopping power formula and its corrections. It includes the most recent advances in the calculations of the various corrections with an extended range of applicability. It is worth mentioning that while all our data for H lie within the CBETHE validity range (E $\geq$ 750 keV), only a few He data lie within the validity range (E $\geq$ 5000 keV).

\section{RESULTS AND DISCUSSION}

    The uncertainty sources considered in this work are summarized in Tab. \ref{tab:uncertainty_sources}, where average values of uncertainties and their respective cumulative contributions to the final uncertainty are presented.
    
    \begin{table*}[h]
    \centering
    \begin{tabular}{lcc}
        \hline
        \textbf{Source of uncertainty} & \textbf{Average} & \textbf{Cumulative} \\
        \textbf{(transmission measurements)} & \textbf{value} & \textbf{contribution} \\
        \hline
        Energy loss measurement & 0.57 & 0.57 \\
        (peak fit, energy and detector calibrations) & & \\
        Thickness determination & 1.0 & 1.15 \\
        Purity of the foil & 1.0 & 1.44 \\
        SRIM data for impurities & 2.0 & 1.5 \\
        \hline
        \textbf{Source of uncertainty } & \textbf{Average} & \textbf{Cumulative} \\
        \textbf{(backscattering measurements)} & \textbf{value} & \textbf{contribution} \\
        \hline
        Energy loss measurement & 1.72 & 1.72 \\
        (edge fit, energy and detector calibrations) & & \\
        Thickness determination & 1 & 2.0 \\
        Purity of the film & - & 2.0 \\
        SRIM data for impurities & - & 2.0 \\
        \hline
    \end{tabular}
    \caption{Uncertainty sources and their respective contributions to the final uncertainty of the stopping cross-section measurements presented in this work. }
    \label{tab:uncertainty_sources}
    \end{table*}

    Other sources of uncertainty, such as geometrical, pulse height defect, and energy calibration, were not considered in this work since they were characterized in previous work and shown to be negligible in our setup.
    
    \begin{figure*}[htb!]
        \centering
        \includegraphics[width=0.85\linewidth]{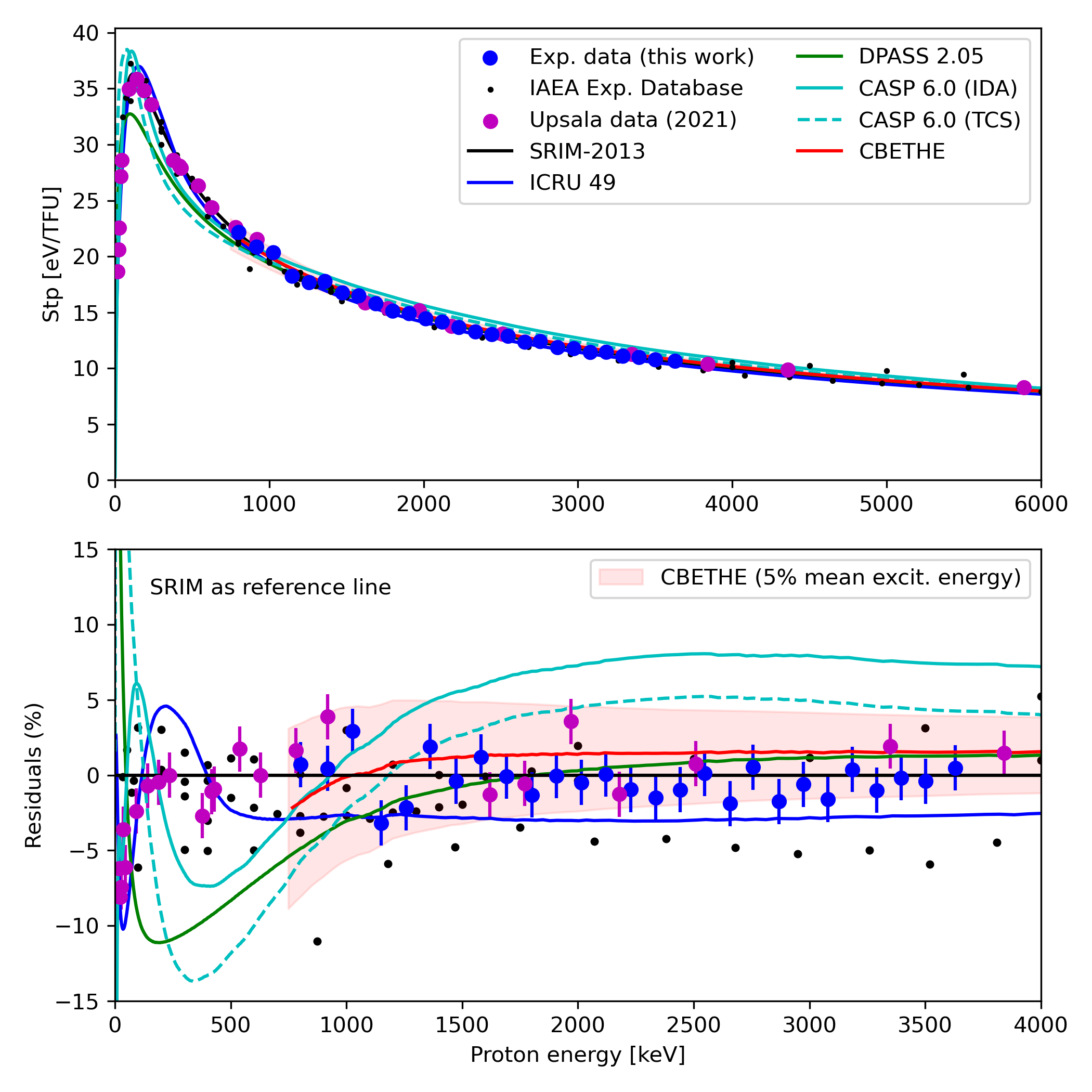}
        \caption{The experimental results for the stopping cross-section of tungsten for protons, obtained in this study using the transmission method, are compared with semi-empirical and theoretical models. Experimental data from the IAEA database \cite{IAEA_expdata} are also included in the comparison, with particular emphasis on recent measurements by the Uppsala group \cite{MORO20211}. 1 TFU = 1$\times$10${}^{-15}$ atoms/cm$^2$. Residuals (bottom) are calculated using SRIM-2013 as a reference. Painted region corresponds to 5\% variation in the mean excitation energy parameter in the Bethe formula.}
        \label{fig:result_trans}
    \end{figure*}

    Fig. \ref{fig:result_trans} presents the experimental values of tungsten stopping power for protons obtained in this work, compared to the values available in the IAEA database, with particular emphasis on recent results from the Uppsala group. The contribution of carbon impurity to stopping was accounted for using SRIM data. The top panel shows the general agreement of the stopping power values together with some model predictions, while the bottom panel provides a closer look at the data by displaying the residuals with respect to the SRIM data. This approach allows for a clearer visualization of fine discrepancies.

    Our data and the Uppsala results agree excellently, with our data providing complementary information by filling energy gaps, such as the range between 1000 up to 1500 keV and 2500 up to 3400 keV. Although the stopping power curve in this energy range is expected to be smooth, our more densely spaced data points help refine the practical uncertainty limits in this region, which we estimate to be approximately 1.5\% root mean square (rms). There are methodological differences between the experimental techniques, and the Uppsala data exhibit slightly higher dispersion.
    
    Regarding the semi-empirical models, we observe that SRIM shows the best agreement with the most recent data above 500 keV (including ours), while ICRU data tend to lie near the lower uncertainty limits in the same energy range. The systematic difference of approximately 2.5\% between SRIM and ICRU for energies greater than 500 keV directly affects the range calculations, with range discrepancies increasing as the energy increases. Experimental data suggest that calculations based on SRIM data tend to be more accurate, with ICRU-based calculations underestimating the range.
    
    CBETHE calculations also show good agreement with the experimental data in the explored energy range, though the expected deviations at lower energies are observed. The validity limit of CBETHE appears to be between 750 and 1000 keV. The shaded band around the CBETHE curve in the residuals plot represents variations in the CBETHE output with a $\pm$5\% variation in the mean excitation energy (in the CBETHE database as 727 eV). The most recent data suggest that this parameter can be better constrained at 709(7) eV.
    
    DPASS also shows good agreement above 1000 keV. Notably, DPASS was expected to align with data within the full energy range, but this agreement was not observed in lower energies within the spread of the experimental data. The CasP exhibits reasonable agreement to energies above 750 keV, with better results obtained with TCS calculation. In energies below 750 keV the agreement is good for the IDA model.

    \begin{figure*}[htb!]
        \centering
        \includegraphics[width=0.85\linewidth]{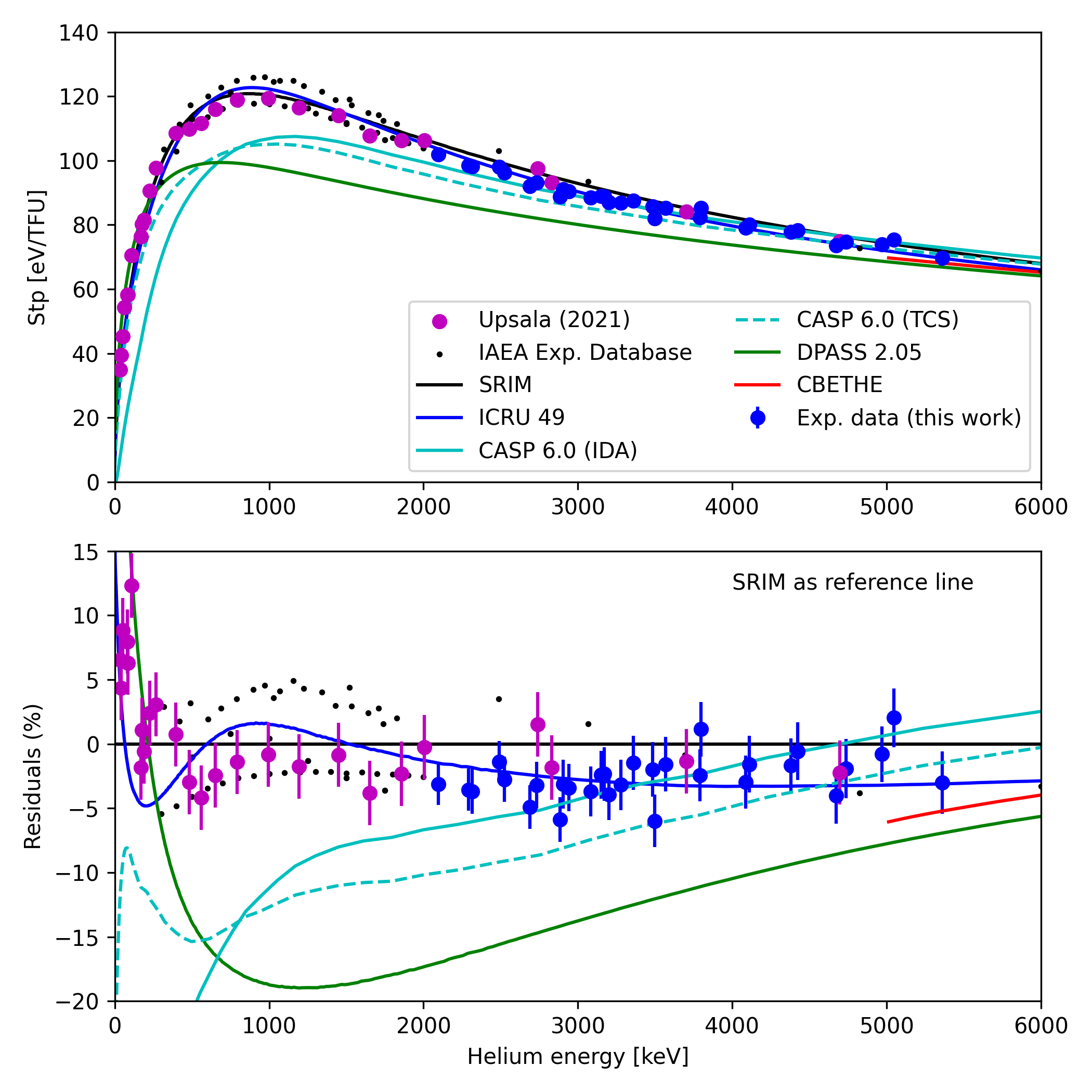}
        \caption{The experimental results for the stopping cross-section of tungsten for helium ions, obtained in this study using the transmission method, are compared with semi-empirical and theoretical models. Experimental data from the IAEA database \cite{IAEA_expdata} are also included in the comparison, with particular emphasis on recent measurements by the Uppsala group \cite{MORO20211}. 1 TFU = 1$\times$10${}^{-15}$ atoms/cm$^2$. Residuals (bottom) are calculated using SRIM-2013 as a reference.}
        \label{fig:result_scat}
    \end{figure*}

    Fig. \ref{fig:result_scat} presents similar comparisons for the obtained values of tungsten stopping power for alpha particles. Once again, we observe the agreement between our data and the Uppsala results. Ours contributes with more densely spaced values, allowing for a refinement of the practical uncertainty limits to approximately 2\% rms.
    
    Due to the wider spread of the experimental data and the larger uncertainty bars, SRIM and ICRU exhibit a similar level of agreement, with ICRU slightly better capturing the fine structure of the experimental data in the residuals plot.
    
    For theoretical models, the overall agreement is not as good as in the proton case. As mentioned above, CBETHE presents good agreement in its limited range of applicability. DPASS, on the other hand, deviates by up to 20\% from the experimental data at the peak of the stopping power curve. CasP shows convergence to the data only at the range limit of higher energies, however, with the adopted option of input parameters, the agreement with the data is similar to DPASS in the energy range of the study. The IDA calculations corrects the TCS towards the experimental data.
    
\section{CONCLUSIONS}

    In this work, we present experimental data on the stopping power of tungsten for protons and alpha particles, with the aim of enhancing the accuracy of ion beam analysis techniques applied to fusion-related materials. Tungsten is widely used in the inner components of fusion reactors, and many studies on light isotope storage and material deposition for this application require greater precision from ion beam-based material characterization methods. The achievement of this improvement depends on increasing confidence in fundamental databases such as those providing stopping power values for tungsten.
    
    Our experimental results and those recently published by the Uppsala group show excellent agreement. In addition to mutual corroboration, our measurements provide a more densely spaced dataset, helping to refine the practical uncertainty limits to approximately 1.5\% for protons and 2\% for alpha particles within the energy range of 1000 and 5000 keV, approximately.
    
    Our work also contributes to the evaluation of the current status of theoretical models to calculate the stopping power of tungsten for light projectiles. For protons, comparisons with experimental data show good agreement in energies above 750 keV, approximately. For alpha particles, we identified divergence between the experimental data and the theoretical models, with theories converging to the Bethe formula. This is supported by the strong agreement between DPASS, CasP and CBETHE above 5000 keV. It is worth mentioning that CasP offers many options for corrections and calculation parameters. In this work, our choice was to keep the model input as homogeneous as possible to the two projectiles under study.  
    
    The semi-empirical models show excellent agreement with the experimental data, remaining within the region defined by data dispersion. SRIM presents better results for protons with the ICRU data lying on the statistical limits of the experiment. However, ICRU data appear to better capture fine differences in the residual analysis. For helium, the situation is the opposite, with SRIM in the statistical limit and ICRU data showing better agreement.

\begin{figure}[htb!]
        \centering
        \includegraphics[width=1.0\linewidth]{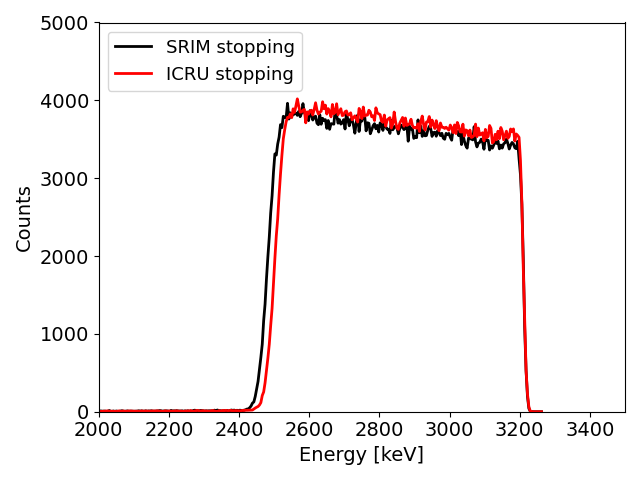}
        \caption{Comparison of RBS simulated spectra of a 630 nm thick tungsten film with SRIM and ICRU stopping cross-section databases. Simulation of 3500 keV helium scattering at 165$^\circ$ scattering angle, with 1 msr solid angle detector and accumulated charge of 10.0 $\mu$C. Noise added artificially with Poisson distribution.}
        \label{fig:conparison}
    \end{figure}

    To illustrate the impact of the observed difference between SRIM and ICRU in the energy interval of 2000 and 3500 keV on a Rutherford backscattering analysis (RBS) for the helium projectile case, we used SIMNRA7 to calculate the RBS spectra of a 630 nm thick tungsten film. The code was fed with both stopping cross sections, and the calculation was performed to a helium particle impinging the film perpendicularly to the surface. The scattering angle was 165$^\circ$, the detector solid angle was 1.0 msr and the accumulated charge was 10.0 $\mu$C. Noise was added adopting a Poisson distribution. The incident energy was 3500 keV. The results are presented in Fig. \ref{fig:conparison}.

    The differences in the two spectra are visible, with the peak width in the calculation that adopts the SRIM database being 3\% larger than the peak in the calculation that adopts the ICRU database (approximately 18 nm in film thickness). Height differences are observable even with the introduction of statistical noise, possibly affecting the determination of relative concentrations. This result illustrates how the depth scale is affected in an NRA or RBS analysis in the region with the largest differences observed in this study. 

\section{Acknowledgment}

    The authors thank the financial support provided by CNPq-INCT-FNA (project number 464898/2014-5). TFS acknowledges the Brazilian funding agency CNPq (project number 406982/2021-0). FM acknowledges CNEN (Project No. 2020.06.IPEN.32). This research used resources from the Laboratory for Materials Analysis with Ion Beams - LAMFI-USP,  of the University of S\~ao Paulo. The authors acknowledge laboratory staff for their assistance during the experiments.

\hfill \break
During the preparation of this work, the authors used ChatGPT by OpenAI to improve language and readability. After using this tool, the authors reviewed and edited the content as needed, and assume full responsibility for the content of the publication.





\bibliographystyle{model1-num-names}
\bibliography{references.bib}








\end{document}